\documentclass[aps, prl, twocolumn, showkeys, floatfix]{revtex4-1}

\usepackage{chemformula} % for chemical symbols and formulae
\usepackage{siunitx} % for si units
\usepackage[inline,shortlabels]{enumitem} % inline numbered lists
\usepackage{hyperref} %hyperref can cause crash if citation is split over two pages. This is remedied by the [draft] option, which can be removed after rearranging the text a bit.
\usepackage{graphicx} % including graphics
\usepackage{tabularx}
\usepackage{threeparttablex} % Tables with notes and citations
\usepackage[section]{placeins}
\usepackage{multirow}
\usepackage{fancyhdr}
\usepackage{dblfloatfix}
\usepackage{float}
\usepackage{caption}
\usepackage{subcaption}
\usepackage{xcolor}
\usepackage{flushend}

\makeatletter
\renewcommand{\fnum@figure}{Fig. \thefigure}
\makeatother

\newcommand{\purple}[1]{\textcolor{purple}{#1}}

\usepackage[justification=raggedright, labelsep=period]{caption}
\begin{document}

    \title{Unusual Ferrimagnetic Ground State in Rhenium Ferrite}
    \author{M. Hussein N. Assadi}
    \email{h.assadi.2008@ieee.org}
    \affiliation{School of Materials Science and Engineering, University of New South Wales, Sydney NSW 2052, Australia.}
    \author{Marco Fronzi}
    \affiliation{College of Engineering, Shibaura Institute of Technology, Toyosu, Koto City, Tokyo 135--8548, Japan.}
    \author{Dorian A. H. Hanaor}
    \affiliation{Fachgebiet Keramische Werkstoffe, Technische Universit\"at Berlin, 10623 Berlin, Germany.}
    \date{2022}
    
    \begin{abstract}
    Through comprehensive density functional calculations, we predict the stability of a rhenium-based ferrite, \ch{ReFe2O4}, in a distorted spinel-based structure. In \ch{ReFe2O4}, all Re and half of the Fe ions occupy the octahedral sites while the remaining Fe ions occupy the tetrahedral sites. All Re ions are predicted to be at a +4 oxidation state with a low spin configuration ($S = 3/2$), while all Fe ions are predicted to be at a +2 oxidation state with a high spin state configuration ($S = 2$). Magnetically, \ch{ReFe2O4} adopts an unconventional ferrimagnetic state in which the magnetic moment of Re opposes the magnetic moments of both tetrahedral and octahedral Fe ions. The spin-orbit coupling is found to cause a slight spin canting of $\sim 1.5^{\mathrm{\circ}}$. The predicted magnetic ground state is unlike the magnetic alignment usually observed in ferrites, where the tetrahedral cations oppose the spin of the octahedral cations. Given that the density of states analysis predicts a half-metallic character driven by the presence of Re $t_{2g}$ states at the Fermi level, this compound shows promise towards potential spintronics applications.
    \end{abstract}

\maketitle

       \section{INTRODUCTION}%\FloatBarrier
Searching for and investigating exotic magnetic phases deepens our fundamental knowledge of complex functional materials and opens new horizons for novel applications [1]. Ferrites are among the most studied magnetic materials, with broad applications in spintronics [2], magnetic data storage [3], magnetically recoverable catalysts [4-7], and microwave guides [8]. The utility of ferrites stems from their high magnetic saturation, high Curie temperatures and controllable coercivity. Ferrites, most commonly synthesised by reactions of \ch{Fe2O3} with a smaller proportion of other metal oxides, encompass a wide range of chemical compositions, stoichiometries, and crystal structures. However, they mostly crystallise into cubic spinel structures, distorted spinels with lower symmetry, [9] or hexagonal structures [10]. Ferrite cations are situated between octahedral and tetrahedral voids, created through the oxygens' closed packed arrangement [11]. Alongside Fe, cation sites may be occupied by Sr, Ba, Pb, or transition metals (TMs). Generally, ferrites are ferrimagnetic where the spin of the tetrahedral cations opposes but does not cancel the spin of the octahedral cations [12]. The second cation's type and its abundance determine the magnetic hardness and saturation of the resultant ferrite, enabling the design of materials with desirable magnetic phase transition temperatures and magnetic coercivity [13].\par
Typically, in complex materials containing multiple magnetic cations, the magnetic ground state is stabilised through competing ferromagnetic and antiferromagnetic superexchange interactions, resulting in frustrated systems with multiple magnetic phase transitions. In ferrites with only fourth row 3d TM ions, these competitions simply stabilise the ferrimagnetic state where the cations on the octahedral site align antiparallel to the cations on the tetrahedral sites [12]. However, in other classes of complex oxides, such as double perovskites, magnetic interactions between 3d and heavier 4d and 5d elements has demonstrated substantially more complex magnetic behaviour [14,15], often deviating from the rules of thumb established by Goodenough and Kanamori [16,17], where exotic magnetic behaviours are often the result of the strong spin-orbit coupling, structural distortions and higher bond covalency between heavier TM ions and oxygen. A detailed review of Goodenough-Kanamori rules can be found in the literature [18,19].\par
With all these exciting developments in perovskite magnetism, one wonders if there are any similar counterparts in ferrites. The idea of harnessing heavier 4d and 5d TM ion ferrites to control anisotropy and magnetostriction in ferrite was proposed by Hansen and Krishnan in 1977 [20]. However, since then, this idea has not attracted the attention it deserves. In the present work, we demonstrate the stability of a rhenium-based ferrite \ch{ReFe2O4} and discusses its unconventional ferrimagnetic ground state through comprehensive density functional calculations.\par

       \section{COMPUTATIONAL SETTINGS}%\FloatBarrier
Spin-polarised collinear and noncollinear density functional calculations were performed with VASP code [21,22], using the projector augmented wave method (PAW) [23] and the Perdew--Burke--Ernzerhof (PBE) exchange-correlation functional [24,25]. To improve the electronic band description, adequate intra-atomic interaction terms ($U_{\mathrm{eff}}$), based on the Liechtenstein \textit{et al.} approach [26], were added to the Fe 3d electrons. The $U$ and $J$ parameters were 3.5 eV and 0.5 eV, respectively, resulting in an effective $U$ ($U_{\mathrm{eff}}$) of 3 eV. Comparable values were reported to improve the band description accuracy of ferrites [27,28]. More specifically a $U_{\mathrm{eff}}$ value of 3 eV is necessary to adequately describe the Fe 3d electrons in oxides [29]. Furthermore, our comprehensive test demonstrated the adequacy of these values (\purple{Fig. S1}). Accurate electronic localisation through the GGA$+U$ formalism is essential for obtaining reliable structures as the atomic forces are sensitive to magnetic moments borne on cations [30]. The energy cut-off was set at 650 eV. The precision key for the rest of the parameters was set \textit{ACCURATE}. The noncollinear calculations were initiated with the \textit{WAVECAR} files calculated with the spin-polarised collinear method to facilitate convergence.\par
To simulate the \ch{ReFe2O4} structure, as shown in Fig. \ref{figure:1}, two Fe ions in the primitive magnetite cell, with the chemical formula \ch{Fe6O8} [31], were substituted with Re ions. As shown in Fig. \ref{figure:1}a--j, we considered all possible rhenium placement scenarios, and for each scenario, we examined various spin alignments, searching for the most stable structure. Substitution at the octahedral sites consistently resulted in lower total energy, so we further investigated all plausible spin alignments in \ch{ReFe2O4} with only octahedral Re. A dense $7 \times 7 \times 7$ k-point mesh, generated with the Monkhorst-Pack scheme of $\sim 0.015$ $\mathrm{\AA}^{-1}$ spacing, consisting of 172 irreducible sampling points in the Brillouin zone, was used for geometry optimisation. For geometry optimisation, the internal coordinates and the lattice parameters were relaxed to energy and force thresholds smaller than $10^{-6}$ eV and 0.02 eV $\mathrm{\AA}^{-1}$ , respectively. No symmetry restriction was applied in geometry optimisation to allow relaxation to lower symmetry, should it be more stable. It is well-known that even the simplest of the spinel ferrites, \ch{Fe3O4}, although cubic ($Fd\bar3m$) at room temperature, transforms to a lower symmetry monoclinic structure ($P2_1/c$) below $\sim 125$ K, through a Verwey phase transition [32].\par
       
       \section{RESULTS AND DISCUSSION}\FloatBarrier

\begin{figure*}[t!]
            \centering
            \includegraphics[width=0.9\linewidth]{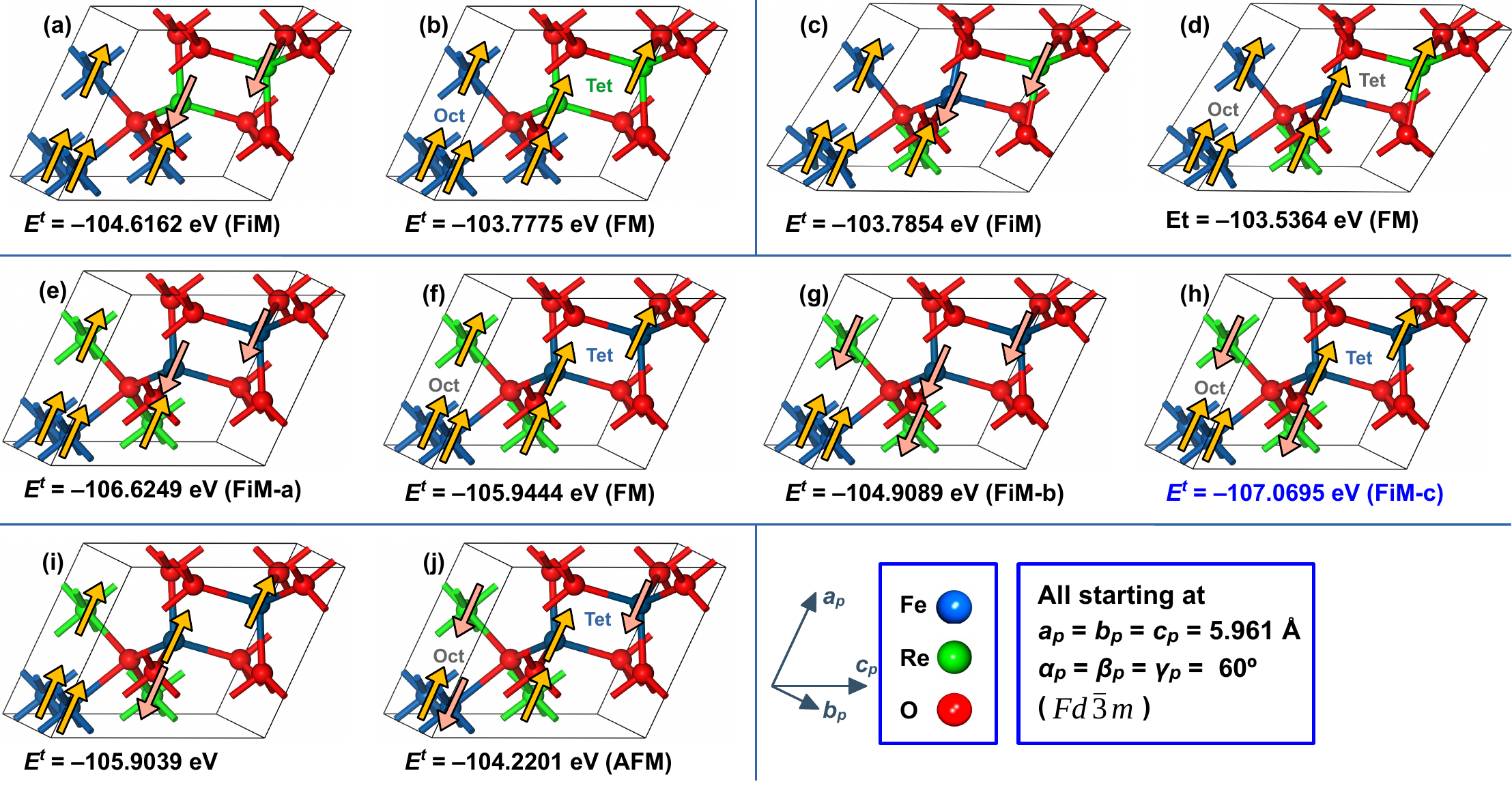}
            \caption{\label{figure:1} The spin configurations used for determining the magnetic ground state of \ch{ReFe2O4}. In configurations \textbf{a} and \textbf{b}, Re ions are at the tetrahedral sites. In \textbf{c} and \textbf{d}, one Re ion is located at the tetrahedral site while the other is located at the octahedral site. In \textbf{e}, \textbf{f}, \textbf{g}, \textbf{h}, \textbf{i}, and \textbf{j}, both Re ions are located at the octahedral sites. The density functional total energy ($E^t$) of each configuration is also shown. FM, FiM, and AFM refer to ferromagnetic, ferrimagnetic and antiferromagnetic, respectively.}
         \end{figure*}

Two cation types based on coordination exist in the spinel ferrite structure: one that is tetrahedrally coordinated and another that is octahedrally coordinated with oxygen ions. The first type represents one-third, while the second represents two-thirds of the total cations in the crystal. We examined three possible rhenium placements to identify the most stable position of the Re ions in \ch{ReFe2O4}. First, both Re ions were placed at the tetrahedral sites while all Fe ions were left at the octahedral site, which is the typical cationic distribution in spinels. Secondly, one Re ion was placed at the tetrahedral site, and the other Re ion was placed at the octahedral site, a configuration usually referred to as an intermediate spinel. Lastly, both Fe ions were placed at tetrahedral sites, while the octahedral sites were equally occupied with Re and Fe, which is referred to as an inverse spinel. We investigated all these cationic distributions by calculating the total energy for two possible ferrimagnetic and ferromagnetic spin alignments. In the ferromagnetic structure, all cations' spin was set parallel, while in the ferrimagnetic structure, the tetrahedral cations' spin was set antiparallel to the spin of the octahedral cations. As shown in Fig. \ref{figure:1}a and b, when Re ions are located at tetrahedral sites, the total energy is relatively high regardless of the direction of the Re spin. However, the ferrimagnetic state is still more stable than the ferromagnetic one. For intermediate rhenium occupancy, as in Fig. \ref{figure:1}c, d, the total energies for both spin alignments were slightly higher than the previous case of complete tetrahedral Re occupation.\par
When Re ions are located at octahedral sites (Fig. \ref{figure:1}e--j), the compound's total energy generally decreased significantly, indicating greater stability. For instance, the aforementioned ferrimagnetic spin alignment of Fig. 1e was more stable than the counterparts with tetrahedral Re (Fig. \ref{figure:1}a) and the mixed Re (Fig. \ref{figure:1}c) configurations by approximately 2 eV per unitcell. The stability of the octahedral Re warranted further investigations of other possible spin alignments of \ch{ReFe2O4} with octahedrally coordinated rhenium. Accordingly, we calculated all different possible ferrimagnetic spin alignments for this ferrite compound with octahedral Re [see configurations (g) and (h)]. In particular, configuration (h) was the most stable among all. In this ferrimagnetic configuration, the Re ions' spin direction is antiparallel to the spins of both octahedral Fe and tetrahedral Fe ions. In this case, the spins of the octahedral Fe and tetrahedral Fe ions were parallel. Configuration (e), with similar spin alignment to a conventional ferrimagnetic inverse spinel such as magnetite, was higher in total energy relative to configuration (h) by 0.4446 eV/u.c. (u.c. is unitcell). Configuration (g), representing the other possible realisation of the ferrimagnetic alignment, also had higher total energy than configuration (h) by 2.1606 eV/u.c. Likewise, the ferromagnetic state in configuration (f) had higher total energy of 1.1251 eV/u.c.\par
To unambiguously confirm the stability of configuration (h), we further calculated the total energy of configuration (i), which is quite similar to configuration (h) except that the spins of the two Re ions were set antiparallel to examine the strength of the magnetic coupling among Re ions. We also examined configuration (j), which is antiferromagnetic; that is, every cation has an antiparallel spin to the one adjacent. Both configurations (i) and (j) had total energies higher than that of configuration (h) by more than 1 eV/u.c. The stability of configuration (h) relative to configurations (i) and (j) demonstrates that pairs of \ch{Re_{Oct}}, \ch{Fe_{Oct}} and \ch{Fe_{Tet}} ions have a strong tendency towards parallel spin alignment among themselves.\par
For the most stable placement of Re which can be expressed as \ch{Fe_{Tet}(ReFe)_{Oct}O4}, configuration (h), which represents the magnetic ground state, is remarkably stable as flipping to even the second most stable spin configuration (e) has an energy cost of 0.4446 eV/u.c. This level of stability is likely to correspond to an ambient Curie temperature in \ch{ReFe2O4}. This prediction is based on a comparison with \ch{CoFe2O4}'s stability margin. A spin-flip from ferrimagnetic to ferromagnetic order in \ch{CoFe2O4} costs 0.536 eV/u.c. [33], resulting in a Curie temperature ($T_C$) of 793 K in bulk cobalt ferrite [34]. Furthermore, the mean-field approximation can estimate the Curie temperature based on the magnetic exchange between Fe and Re sublattice systems as follows [35]:
          \begin{equation}
           \frac{3}{2} k_B T_C = \sum_{i \neq j} \textbf{\textit{J}}_{ij},
          \end{equation}
in which $k_B$ is the Boltzmann constant and $\textbf{\textit{J}}_{ij}$ is the pair exchange coupling parameter between sites $i$ and $j$. Assuming the nearest neighbour interaction between Re and Fe to be the most significant, Equation 1 yields $T_C = 734.9$ K, which is close to the comparison made earlier.\par

\begin{figure*}[t!]
            \centering
            \includegraphics[width=0.9\linewidth]{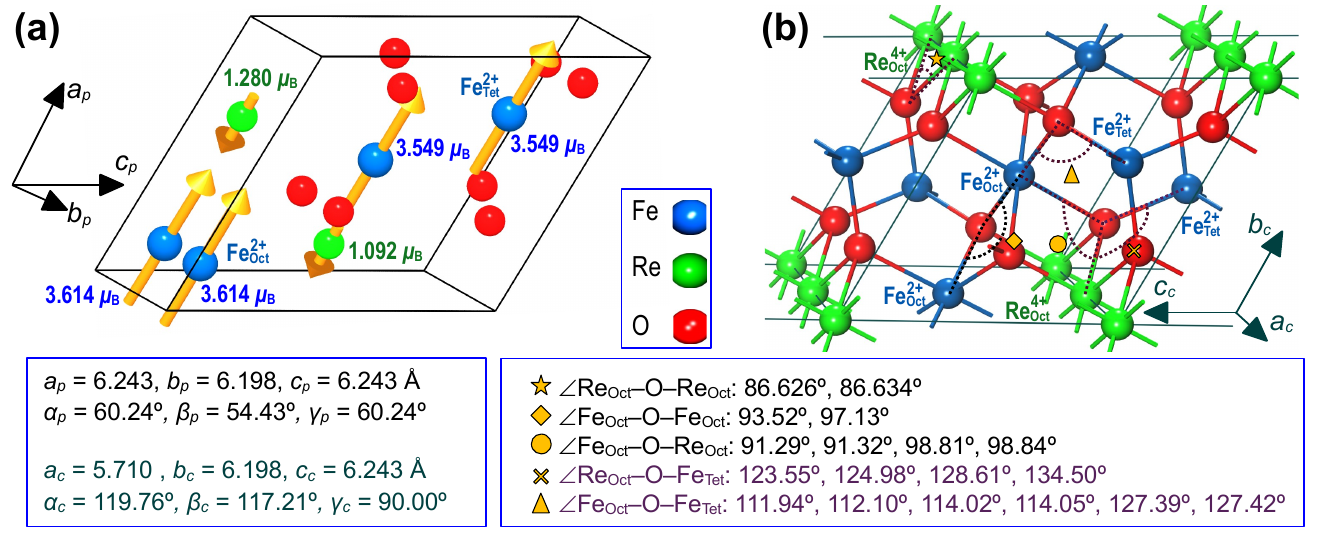}
            \caption{\label{figure:2} \textbf{a} The optimised primitive cell of \ch{ReFe2O4} with SOC taken into account, along with the magnetic moments borne on all cations. \textbf{b} The conventional representation of the optimised \ch{ReFe2O4}, which has a triclinic symmetry. In \textbf{b}, all possible bond angles between cations are also presented. The corresponding yellow marks show a representative of each given angle. The lattice parameters shown are either indexed with \textit{p} or \textit{c}, indicating the primitive and conventional cell dimensions, respectively.}
         \end{figure*}

Given that Re is a relatively heavier sixth-row transition element, we anticipate that the role of spin-orbit coupling (SOC) is potentially significant in \ch{ReFe2O4} [36]. To examine the significance of SOC, we re-optimised the most stable configuration (h) with spin-orbit coupling taken into consideration. The structural relaxation was minor as none of the internal coordinates and the lattice parameters changed by more than $\sim 1\%$. However, the total energy was lowered to $-107.7975$ eV/u.c., indicating that SOC accounts for $\sim 0.67\%$ of the total energy. Similarly, SOC is not anticipated to change the magnetic alignment in \ch{ReFe2O4} as other competing configurations are also estimated to have their total energy lowered by approximately the same amount when SOC is considered (\purple{Table S1}). SOC brings about magnetic noncollinearity by coupling the spin to the orbital degrees of freedom, of which the latter depends on the lattice environment. The noncollinear magnetic alignment of \ch{ReFe2O4} is shown in Fig. \ref{figure:2}a. More precisely, the spins of the Re ions formed a tight angle of $1.329^{\circ}$ between each other, \textit{i.e.}, nearly parallel; and the net spin of the Re ions formed an angle of $177.593^{\circ}$ with the net spin of the Fe ions, \textit{i.e.}, nearly antiparallel. It can be seen that the degree of noncollinearity is relatively small as the spin directions are to a great extent similar to that of the collinear alignment of configuration (h). The net magnetisation of the whole compound was calculated to be 6.090 $\mu_{\mathrm{B}}$/f.u. (f.u. stands for formula unit), which is approximately 1.5 times larger than that of the saturation magnetic moment of magnetite [37]. A more qualitative description of the effect of the $U_{\mathrm{eff}}$ choice on the total magnetisation and the local magnetisation of the Re and Fe ions in this configuration is given in \purple{Table S2}.\par
                    
\begin{figure}[tb]
            \centering
            \includegraphics[width=1.0\linewidth]{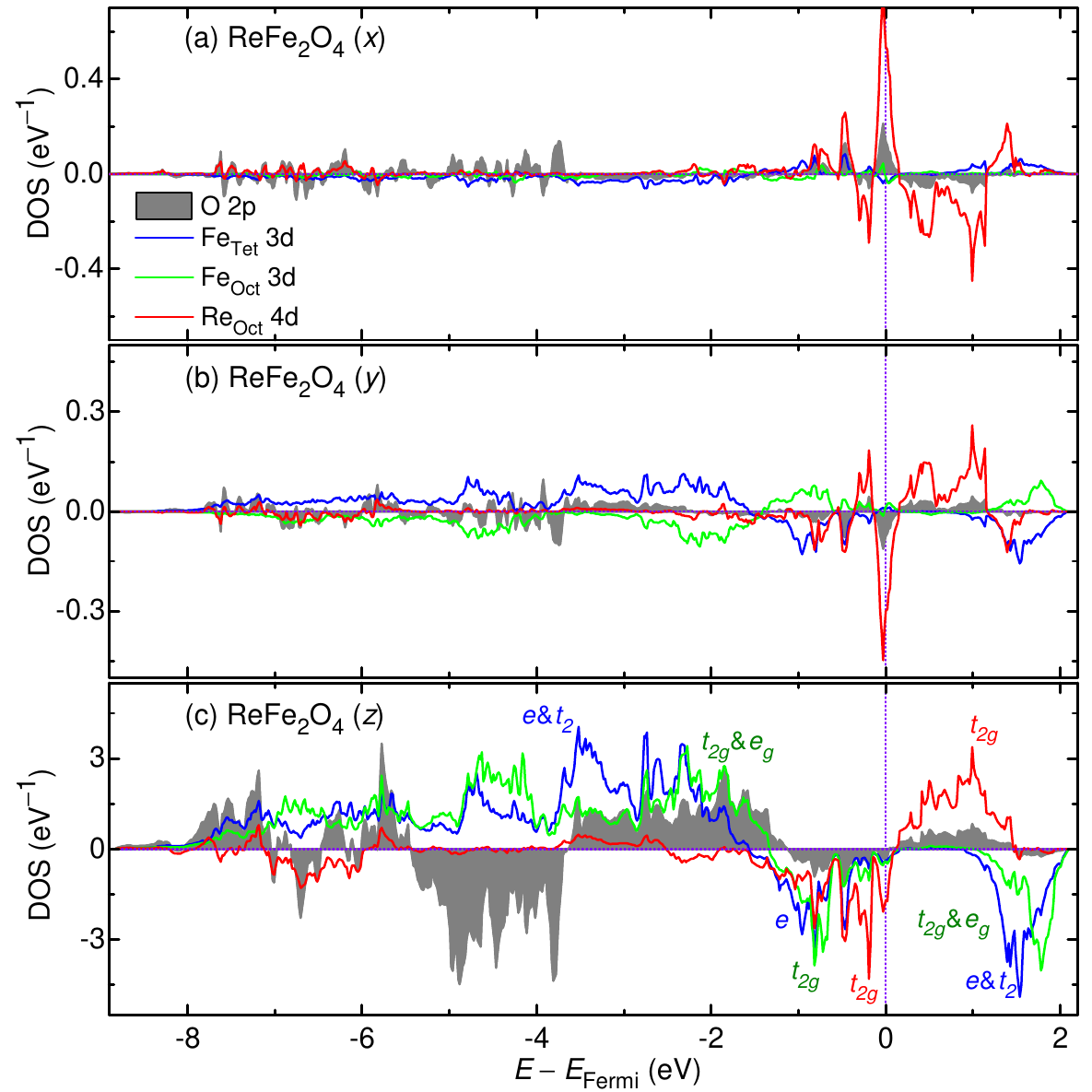}
            \caption{\label{figure:3} Partial density of states of \ch{ReFe2O4} at its most stable magnetic state [configuration (\textbf{h}) of Fig. \ref{figure:1}], calculated with spin-orbital coupling considered. The density of states is projected along an orthogonal frame having the $x$, $y$, and $z$ axes. The $z$ axis of this frame coincides along the lattice parameter $c_p$ of the primitive lattice parameter, shown in the lower row of Fig. \ref{figure:1}.}
         \end{figure}

The density of states (DOS) in \ch{ReFe2O4}, calculated considering SOC, are presented in Fig. \ref{figure:3}. Here, DOS is projected onto the axes of an orthogonal frame of which the $z$ axis is parallel to the $c_p$ direction of the \ch{ReFe2O4} primitive cell. First, we notice that the DOS magnitude along the $x$ and $y$ axes is $\sim 3\%$ of the DOS magnitude along the $z$ axis, indicating a slight deviation from linearity and corroborating the magnetic moment orientations obtained in Fig. \ref{figure:2}a. Along the $z$ axis, (Fig. \ref{figure:3}c), we can see that both \ch{Fe_{Tet}} and \ch{Fe_{Oct}} are in high-spin and +2 oxidisation states as one spin channel, comprising of five electrons per Fe, for both \ch{Fe_{Tet}} and \ch{Fe_{Oct}}, is fully occupied while the other spin channel is only partly occupied. Therefore, the electronic configuration for \ch{Fe_{Tet}^{2+}} is $e_2$ $\uparrow$ $t_2^3$ $\uparrow$ $e^1$ $\downarrow$, while for \ch{Fe_{Oct}^{2+}}, the electronic configuration is $t_{2g}^3$ $\uparrow$ $e_g^2$ $\uparrow$ $t_{2g}^1$ $\downarrow$. Moreover, Re is in low-spin state with +4 oxidation. According to its partial DOS, Re has a fully occupied $t_{2g}$ $\uparrow$ states, which are immediately followed by its empty $t_{2g}$ $\downarrow$ states. The Fermi level crosses the tail of $t_{2g}$ $\uparrow$ states, giving rise to half-metallic conduction, as no states are available at the Fermi level with opposing net spin direction. Moreover, because of the larger crystal field acting on Re's 5d states, Re's empty $e_g$ states are located at $\sim 4$ eV above the Fermi level (\purple{Fig. S2}). The \ch{ReFe2O4} half-metallicity can be utilised for magneto-resistive response [38,39] or near-perfect spin-polarised current injection [40,41].\par
As shown in Fig. \ref{figure:3}, the Fe 3d states are spread through the conduction band, hybridising extensively with O, while Re 5d states are mainly concentrated within 2 eV below the Fermi level. Nonetheless, in the region of $-2 < E-E_{\mathrm{Fermi}} < 0$, the spin-down Re and Fe states and O 2p states hybridise together, facilitating the magnetic superexchange interactions that stabilise the magnetic ground state. Furthermore, the net magnetic moments borne on all cations, as shown in Fig. \ref{figure:2}, are smaller than ideal ions. For high spin \ch{Fe^{2+}} (3d$^6$), either in tetrahedral or octahedral coordination, the magnetic moment should have been $4 \mu_{\mathrm{B}}$. For octahedral \ch{Re^{4+}} (5d$^3$), the magnetic moment should have been $3 \mu_{\mathrm{B}}$. However, since Fe--O or Re--O bonds are not purely ionic but possess a degree of covalency, the magnetisation of transition metal ions is expected to be lower than the purely ionic values [42]. The reduction in magnetisation is more profound in heavier transition metal ions are their bonds are more covalent [33]. A more quantitative description of the electronic localisation function is provided in \purple{Fig. S3}.\par

As shown in Fig. \ref{figure:2}a, the primitive cell of \ch{ReFe2O4} is substantially transformed by geometry optimisation. The lattice parameters of the optimised primitive cell do not conform to the high symmetry of the initial structure as its lattice parameters asymmetrically changed and its volume expanded. The initial structure volume, which was based on magnetite, was 155.225 $\mathrm{\AA}^3$, while the optimised structure had a volume of 163.025 $\mathrm{\AA}^3$. The ionic radius of Fe can explain the expansion of the volume upon Re substitution at the tetrahedral site. The radius of high-spin \ch{Fe_{Tet}^{2+}} in \ch{ReFe2O4} is 0.63 $\mathrm{\AA}$. In magnetite, the tetrahedral site is occupied by \ch{Fe^{3+}} with a smaller radius of 0.49 $\mathrm{\AA}$. The \ch{Re^{4+}} radius in octahedral coordination is 0.63 $\mathrm{\AA}$ which is quite close to the replaced \ch{Fe^{3+}} radius (0.65 $\mathrm{\AA}$) and is not expected to be a substantial drive in the structural transformation.\par
We examined the optimised \ch{ReFe2O4} primitive cell's symmetry to investigate the magnetic exchange among all cations. A triclinic symmetry was detected through the FINDSYM symmetry detection algorithm [43] with a tight tolerance of 0.00001 $\mathrm{\AA}$ for lattice parameters (CIF provided in the supplementary information). The conventional cell with triclinic symmetry is shown in Fig. \ref{figure:2}b. Detecting all possible magnetic exchanges that stabilise the predicted ground state magnetism---configuration (h) re-optimised with SOC considered--is easier in the symmetry-imposed structure. The TM--O--TM bond angles for this structure are all listed in Fig. \ref{figure:2}. The TM--O--TM bond angles between cations on tetrahedral and octahedral sites are obtuse and thus dominate the magnetic exchange interactions, as the superexchange interaction magnitude is proportional to $cos^2$($\angle$TM--O--TM). The higher total energy of configuration (g) indicates the superexchange between \ch{Re_{Oct}^{4+}} and \ch{Fe_{Tet}^{2+}} is antiferromagnetic, while the higher total energy of configuration (e) indicates that the superexchange between \ch{Fe_{Oct}^{2+}} and \ch{Fe_{Tet}^{2+}} is ferromagnetic. The TM--O--TM bonds among tetrahedral sites are all nearly right angles, indicating a minimal orbital overlap favouring weaker ferromagnetic superexchange. The strength of this ferromagnetic exchange can be estimated from the total energy of configurations (g) and (h) of Fig. \ref{figure:1}, showing that setting adjacent cations to antiferromagnetic coupling raises the total energy.\par    
Finally, we examine the stability of \ch{ReFe2O4} in competing metallic and oxide phases. The compound's formation enthalpy ($ \Delta H_{\mathrm{metallic}}$) was calculated relative to the metallic Re (hexagonal paramagnetic) and Fe (body-centred ferromagnetic) phases, and gaseous \ch{O2} was calculated as
         \begin{equation}
          \Delta H_{\mathrm{metallic}} = E^t(\ch{ReFe2O4}) - E^t(\ch{Re}) - 2E^t(\ch{Fe}) - 2E^t(\ch{O2}).
         \end{equation}
Here, $E^t$ is the density functional total energy.  $\Delta H_{\mathrm{metallic}}$ was found to be $-5.1924$ eV/f.u. The formation enthalpy relative to the competing oxide phase ( $\Delta H_{\mathrm{oxide}}$) was calculated as
          \begin{equation}
            \Delta H_{\mathrm{oxide}} =  E^t(\ch{ReFe2O4}) - E^t(\ch{ReO2}) – 2E^t(\ch{FeO}).
          \end{equation}
Here, \ch{ReO2} was the most stable \ch{Re^{4+}} oxide in orthorhombic structure (materials project identifier mp-7228 [44]), and \ch{FeO} was the most stable \ch{Fe^{2+}} oxide in monoclinic structure (materials project identifier mp-1279742 [44]). $\Delta H_{\mathrm{oxide}}$ was found to be $-1.1903$ eV/f.u. Given the negative  $\Delta H$ values, we can conclude that \ch{ReFe2O4} is stable against decomposition to oxides of its constituent elements and \ch{Re^{4+}} and \ch{Fe^{2+}}. For the future synthesis of \ch{ReFe2O4}, one can draw inspiration from the recently developed green fabrication methods for ferrites [45].\par

       \section{CONCLUSIONS}\FloatBarrier
Using density functional calculations, considering spin-orbit coupling, we predict that a Re-based ferrite \ch{ReFe2O4} is stable in a distorted spinel structure with reduced triclinic symmetry ($P\bar1$), and adopts an unconventional magnetic ordering in where Re spin opposes the spin of both \ch{Fe_{Tet}} and \ch{Fe_{Oct}}, while \ch{Fe_{Tet}} and \ch{Fe_{Oct}} have parallel spin alignment among themselves. The net magnetic moment of this compound is evaluated at 6.090 $\mu\mathrm{_B/f.u.}$ which is about 1.5 times greater than that of magnetite. The magnetic ground state is remarkably stable as flipping any spin incurs an energetic cost of at least 0.2223 eV/f.u., which is likely to correspond to an ambient Curie temperature. The compound is predicted to be half-metallic, which implies that this compound may be useful towards applications in spintronics where the spin polarisation of conduction electrons is desired.\par
             
       \section{CONFLICTS OF INTEREST}
The authors declare that there is no conflict of interest.\par
       
      \section{ACKNOWLEDGMENTS}
The authors gratefully acknowledge the funding of this project by computing time provided by the Paderborn Center for Parallel Computing (PC$^2$).\par

       \section{VERSION OF RECORD}
M. Hussein N. Assadi, Marco Fronzi and Dorian A. H. Hanaor \underline{Unusual ferrimagnetic ground state in rhenium ferrite} \textit{Eur. Phys. J. Plus} (2022) \textbf{137}, 21. \url{https://doi.org/10.1140/epjp/s13360-021-02277-z}\par

       \section{REFERENCES}
 [1] A. Hirohata, K. Yamada, Y. Nakatani, I.-L. Prejbeanu, B. Dieny, P. Pirro, and B. Hillebrands, J. Magn. Magn. Mater. 509, 166711 (2020). \url{https://doi.org/10.1016/j.jmmm.2020.166711}\par
 [2] R. K. Kotnala and J. Shah, in Handbook of Magnetic Materials, edited by K. H. J. Buschow (Elsevier, 2015), Vol. 23, pp. 291. \url{https://doi.org/10.1016/B978-0-444-63528-0.00004-8}\par
 [3] J. C. Mallinson, The foundations of magnetic recording. (Academic Press, 2012). \url{https://doi.org/10.1016/B978-0-12-466626-9.50008-3}\par
 [4] E. Casbeer, V. K. Sharma, and X.-Z. Li, Sep. Purif. Technol. 87, 1 (2012). \url{https://doi.org/10.1016/j.seppur.2011.11.034}\par
 [5] E. Doustkhah, M. Heidarizadeh, S. Rostamnia, A. Hassankhani, B. Kazemi, and X. Liu, Mater. Lett. 216, 139--143 (2018). \url{https://doi.org/10.1016/j.matlet.2018.01.014}\par
 [6] E. Doustkhah and S. Rostamnia, J. Colloid Interface Sci. 478, 280--287 (2016). \url{https://doi.org/10.1016/j.jcis.2016.06.020}\par
 [7] S. Rostamnia and E. Doustkhah, J. Magn. Magn. Mater. 386, 111--116 (2015). \url{https://doi.org/10.1016/j.jmmm.2015.03.064}\par
 [8] J. D. Adam, L. E. Davis, G. F. Dionne, E. F. Schloemann, and S. N. Stitzer, IEEE Trans. Microw. Theory Tech. 50 (3), 721 (2002). \url{https://doi.org/10.1109/22.989957}\par
 [9] R. Valenzuela, Phys. Res. Int. 2012, 591839 (2012). \url{https://doi.org/10.1155/2012/591839}\par
[10] R. C. Pullar, Prog. Mater. Sci. 57 (7), 1191 (2012). \url{https://doi.org/10.1016/j.pmatsci.2012.04.001}\par
[11] J. L. Dormann and M. Nogues, J. Phys. Condens. Matter 2 (5), 1223 (1990). \url{https://doi.org/10.1088/0953-8984/2/5/014}\par
[12] U. Ozgur, Y. Alivov, and H. Morkoc, J. Mater. Sci. Mater. Electron. 20 (9), 789 (2009). \url{https://doi.org/10.1007/s10854-009-9923-2}\par
[13] V. G. Harris and A. S. Sokolov, J. Supercond. Nov. Magn. 32 (1), 97--108 (2019). \url{https://doi.org/10.1007/s10948-018-4928-9}\par
[14] A. A. Aczel, P. J. Baker, D. E. Bugaris, J. Yeon, H. C. zur Loye, T. Guidi, and D. T. Adroja, Phys. Rev. Lett. 112 (11), 117603 (2014). \url{https://doi.org/10.1103/PhysRevLett.112.117603}\par
[15] S. Fuchs, T. Dey, G. Aslan-Cansever, A. Maljuk, S. Wurmehl, B. Buchner, and V. Kataev, Phys. Rev. Lett. 120 (23), 237204 (2018). \url{https://doi.org/10.1103/PhysRevLett.120.237204}\par
[16] N. S. Rogado, J. Li, A. W. Sleight, and M. A. Subramanian, Adv. Mater. 17 (18), 2225--2227 (2005). \url{https://doi.org/10.1002/adma.200500737}\par
[17] R. Morrow, K. Samanta, T. Saha Dasgupta, J. Xiong, J. W. Freeland, D. Haskel, and P. M. Woodward, Chem. Mater. 28 (11), 3666--3675 (2016). \url{https://doi.org/10.1021/acs.chemmater.6b00254}\par
[18] S. V. Streltsov and D. I. Khomskii, Phys. Usp. 60 (11), 1121--1146 (2017). \url{https://doi.org/10.3367/UFNe.2017.08.038196}\par
[19] A. Bencini and D. Gatteschi, in Electron Paramagnetic Resonance of Exchange Coupled Systems (Springer, Berlin, Germany, 1990), pp. 1--19. \url{https://doi.org/10.1007/978-3-642-74599-7}\par
[20] P. Hansen and R. Krishnan, J. phys. Colloq. 38 (C1), C1147--C1155 (1977). \url{https://doi.org/10.1051/jphyscol:1977130}\par
[21] G. Kresse and J. Furthmuller, Comput. Mater. Sci. 6 (1), 15 (1996). \url{https://doi.org/10.1016/0927-0256(96)00008-0}\par
[22] G. Kresse and J. Furthmuller, Phys. Rev. B 54 (16), 11169 (1996). \url{https://doi.org/10.1103/PhysRevB.54.11169}\par
[23] J. Sun, M. Marsman, G. I. Csonka, A. Ruzsinszky, P. Hao, Y.-S. Kim, G. Kresse, and J. P. Perdew, Phys. Rev. B 84 (3), 035117 (2011). \url{https://link.aps.org/doi/10.1103/PhysRevB.84.035117}\par
[24] J. P. Perdew, K. Burke, and M. Ernzerhof, Phys. Rev. Lett. 77 (18), 3865 (1996). \url{https://doi.org/10.1103/PhysRevLett.77.3865}\par
[25] J. P. Perdew, K. Burke, and M. Ernzerhof, Phys. Rev. Lett. 78 (7), 1396 (1997). \url{https://doi.org/10.1103/PhysRevLett.78.1396}\par
[26] A. I. Liechtenstein, V. I. Anisimov, and J. Zaanen, Phys. Rev. B 52 (8), R5467 (1995). \url{https://doi.org/10.1103/PhysRevB.52.R5467}\par
[27] M. Derzsi, P. Piekarz, P. T. Jochym, J. Łażewski, M. Sternik, A. M. Oles, and K. Parlinski, Phys. Rev. B 79 (20), 205105 (2009). \url{https://doi.org/10.1103/PhysRevB.79.205105}\par
[28] G. Marschick, J. Schell, B. Stoger, J. Goncalves, M. O. Karabasov, D. Zyabkin, A. Welker, D. Gartner, I. Efe, and R. Santos, Phys. Rev. B 102 (22), 224110 (2020). \url{https://doi.org/10.1103/PhysRevB.102.224110}\par
[29] Y. Meng, X.-W. Liu, C.-F. Huo, W.-P. Guo, D.-B. Cao, Q. Peng, A. Dearden, X. Gonze, Y. Yang, J. Wang, H. Jiao, Y. Li, and X.-D. Wen, J. Chem. Theory Comput. 12 (10), 5132--5144 (2016). \url{https://doi.org/10.1021/acs.jctc.6b00640}\par
[30] A. Pham, M. H. N. Assadi, A. B. Yu, and S. Li, Phys. Rev. B 89 (15), 155110 (2014). \url{https://doi.org/10.1103/PhysRevB.89.155110}\par
[31] M. H. N. Assadi, J. J. Gutierrez Moreno, and M. Fronzi, ACS Appl. Energy Mater. 3 (6), 5666 (2020). \url{https://doi.org/10.1021/acsaem.0c00640}\par
[32] F. Walz, J. Phys. Condens. Matter 14 (12), R285 (2002). \url{https://doi.org/10.1088/0953-8984/14/12/203}\par
[33] M. H. N. Assadi and H. Katayama-Yoshida, J. Phys. Soc. Jpn. 88 (4), 044706 (2019). \url{https://doi.org/10.7566/JPSJ.88.044706}\par
[34] D. S. Mathew and R.-S. Juang, Chem. Eng. J. 129 (1), 51--65 (2007). \url{https://doi.org/10.1016/j.cej.2006.11.001}\par
[35] P. W. Anderson, in Solid State Physics, edited by F. Seitz and D. Turnbull (Academic Press, 1963), Vol. 14, pp. 99. \url{https://doi.org/10.1016/S0081-1947(08)60260-X}\par
[36] M. H. N. Assadi, J. J. Gutierrez Moreno, D. A. H. Hanaor, and H. Katayama-Yoshida, Phys. Chem. Chem. Phys. 23, 20129--20137 (2021). \url{https://doi.org/10.1039/D1CP02164H}\par
[37] S. K. Banerjee and B. M. Moskowitz, in Magnetite biomineralization and magnetoreception in organisms, edited by J. L. Kirschvink, D. S. Jones, and B. J. MacFadden (Springer, Boston, MA, 1985), Vol. 5, pp. 17--41. \url{https://doi.org/10.1007/978-1-4613-0313-8}\par
[38] A. Bratkovsky, Phys. Rev. B 56 (5), 2344--2347 (1997). \url{https://doi.org/10.1103/PhysRevB.56.2344}\par
[39] A. M. Haghiri-Gosnet, T. Arnal, R. Soulimane, M. Koubaa, and J. P. Renard, Phys. Status Solidi A 201 (7), 1392--1397 (2004). \url{https://doi.org/10.1002/pssa.200304403}\par
[40] C. M. Fang, G. A. d. Wijs, and R. A. d. Groot, J. Appl. Phys. 91 (10), \url{8340--8344 (2002). https://doi.org/10.1063/1.1452238}\par
[41] X. Hu, Adv. Mater. 24 (2), 294--298 (2012). \url{https://doi.org/10.1002/adma.201102555}\par
[42] S. V. Streltsov and D. I. Khomskii, Proc. Nat. Acad. Sci. USA 113 (38), 10491 (2016). \url{https://doi.org/10.1073/pnas.1606367113}\par
43	H. T. Stokes and D. M. Hatch, J. Appl. Crystal. 38 (1), 237 (2005). \url{https://doi.org/10.1107/S0021889804031528}\par
[44] A. Jain, S. P. Ong, G. Hautier, W. Chen, W. D. Richards, S. Dacek, S. Cholia, D. Gunter, D. Skinner, G. Ceder, and K. A. Persson, APL Mater. 1 (1), 011002 (2013). \url{https://doi.org/10.1063/1.4812323}\par
[45] S. E. Shirsath, X. Liu, M. H. N. Assadi, A. Younis, Y. Yasukawa, S. K. Karan, J. Zhang, J. Kim, D. Wang, A. Morisako, Y. Yamauchi, and S. Li, Nanoscale Horiz. 4 (2), 434 (2019). \url{http://dx.doi.org/10.1039/C8NH00278A}\par

\end{document}